# Combination effect of growth enhancers and carbon sources on synthesis of single-walled carbon nanotubes from solid carbon growth seeds

*Mengyue Wang, *[a] Yuanjia Liu, [a] Manaka Maekawa,[a] Michiharu Arifuku, [b] Noriko Kiyoyanagi, [b] Taiki Inoue, [a] Yoshihiro Kobayashi*[a]*

a Department of Applied Physics, Osaka University, Suita, Osaka 565-0871, Japan

b Nippon Kayaku Co., Ltd., 31-12, Shimo 3-chome, Kita-ku, Tokyo 115-8588, Japan

*Email: wang.my@ap.eng.osaka-u.ac.jp, kobayashi@ap.eng.osaka-u.ac.jp

**Abstract:**

In the synthesis of highly crystalline single-walled carbon nanotubes (SWCNTs), high growth temperatures are preferred, while the formation of impurity amorphous carbon (a-C) causes the termination of SWCNT growth and degrades its properties. To remove this by-product, $H_2O$ and $CO_2$ have been employed in metal-catalyzed SWCNT growth systems because of their oxidizing ability. Recently, nonmetallic nanoparticles have become one of the growth seed candidates because of their high melting points and fewer metal impurities. In this study, by using nanodiamond-derived carbon nanoparticles as the solid growth seeds, we investigated the effect of $CO_2$ and $H_2O$ on high-temperature SWCNT growth with two types of carbon-source supplies: $C_2H_2$ and $C_2H_4$. In this growth system, $H_2O$ showed oxidizing ability to etch a-C with either carbon sources. $CO_2$ exhibited a similar a-C formation-preventing role in $C_2H_4$-supplied growth and achieved higher-purity SWCNTs with higher concentration of $C_2H_4$ than the case of $H_2O$. However, in contrast to the other combinations, $CO_2$ injection in the $C_2H_2$-supplied growth significantly enhanced the formation of a-C rather than the removal of it while the yield of SWCNTs was also increased, indicating the occurrence of the dehydration reaction between $CO_2$ and $C_2H_2$. The present findings will lead to efficient growth of high-quality SWCNTs from nonmetallic growth seeds with the use of growth enhancers.

# 1. Introduction

Since their discovery in 1991, carbon nanotubes (CNTs) [1] have been a subject of great interest because of their exceptional properties, which present high potential in electronic [2], thermal [3], and mechanical applications [4]. Further, with the development of experimental techniques, both single-walled CNTs (SWCNTs) [5, 6] and multiwalled CNTs (MWCNTs) [7] have been obtained controllably. In particular, SWCNTs have received wide attention in the application of flexible electronic devices [8], chemical sensors [9], field-effect transistors (FETs) [10], and quantum light sources [11] owing to their superior properties such as high carrier mobility, structure-dependent electronic band structure, and high sensitivity to the surface environment [12]. Various methods have been developed to obtain SWCNTs, e.g., arc discharge [13], laser ablation [14], and chemical vapor deposition (CVD) [15]. Nevertheless, limited by the uncontrollability of SWCNT structures in terms of crystallinity, chirality, length, etc., experimentally synthesized SWCNTs tend to exhibit lower performances than expected [16-18]. The crystallinity of SWCNTs is highly related to mechanical strength [19] as well as thermal [20] and electrical conductivity [21-23]; accordingly, it should be one of the most important characteristics to improve in the SWCNT growth process. The degradation of crystallinity and subsequently the properties occurs through the formation of lattice defects, including the pentagon–heptagon pair (5–7) [24, 25], the pentagon–heptagon–heptagon–pentagon (5–7–7–5) defects (often called as Stone-Wales defects) [26], ad-dimer [27, 28], and vacancy [29].

To achieve highly crystalline SWCNTs, a high temperature has been recommended for the CVD process because it assists SWCNTs in overcoming the activation energy of defect healing during growth [24, 30]. Along with the increase in growth temperature, the selection of growth seeds that possess high-temperature stability is a necessary factor for the synthesis of highly crystalline SWCNTs. Unfortunately, traditional metal catalysts, such as iron [31, 32], nickel [33], cobalt [34], and related alloys [35], cannot avoid aggregation because of their low melting point [36, 37]. Moreover, even though they could be used as solid catalysts, e.g., W [38], Mo [38], Re [38], and TiC [39], metal nanoparticles are regarded as an impurity in some applications and must be removed through post-treatment procedures, which causes damage to the as-grown

SWCNTs [40]. Nonmetallic nanoparticles, such as Si [41], SiC [41], Ge [41], and nanodiamond (ND) [42, 43], which possess desirable high melting points, have gradually attracted attention as alternative growth seeds of CNTs. Among these growth seeds, ND [42] possesses the notable advantage of avoiding the introduction of unnecessary elements owing to its composition, i.e., carbon.

Another important point to be considered in the CVD procedure is the deposition of amorphous carbon (a-C). As a byproduct formed during SWCNT growth, a-C is adsorbed on the surface of growth seeds and the nanotube sidewall. This results in blocking of the active sites of the seeds and shortening of growth lifetime, which decreases the synthesis efficiency [44]. In a metal-catalyzed CNT growth system, to prevent the formation of a-C, some types of oxidants were employed as growth enhancers to etch away a-C. Water ($H_2O$)-assisted CVD methods helped realize the growth of millimeter-long SWCNTs by maintaining the activity of the metal catalyst by selectively removing the deposited a-C on its surface [45]. In addition, as the etchant, $H_2O$ exhibited its role in CNT growth kinetics [46, 47]. However, the strong oxidation activity of $H_2O$ limits the supply of the enhancer to the ppm level, while a low concentration results in loss of uniformity of the enhancer in a large-scale reactor [48]. Further, a high concentration of $H_2O$ leads to the possibility of etching of both a-C and CNTs [49]. To date, carbon dioxide ($CO_2$)—another known growth enhancer with less oxidative activity—has been employed with a higher concentration to induce oxidation reactions [50-58]. A mild etching effect is the advantage of $CO_2$ in balancing a-C etching and the preservation of CNTs from damage during the mass production of CNTs.

Recently, other enhancing roles of $CO_2$ in metal-catalyzed CNT growth processes have been found when using acetylene ($C_2H_2$) [53, 59] or methane ($CH_4$) [60] as the carbon feedstock. One of its roles is to promote the dehydrogenation reaction of hydrocarbons, which are used as a carbon source in CNT growth. For synthesizing CNTs via CVD, the carbon source is flown into the reactor and the deposition process proceeds with pyrolysis reactions—including the dehydrogenation reaction—to form the intermediate products, which adsorb on the surface of catalysts or growth seeds and are used to construct the CNT structure. When a carbon source such as $C_2H_2$ was mixed with $CO_2$ in an equimolar ratio, a dramatic increase in the CNT growth yield and initial growth rate was observed [53]. To prove the occurrence of the $CO_2$-promoted hydrocarbon dehydrogenation reaction, further evidence was provided through a theoretical calculation. When $C_2H_2$ was taken as the carbon source, with the injection of $CO_2$, the calculated energy barrier of the dissociation of the first H atom from $C_2H_2$ decreased [59].

Despite progress in the use of growth enhancers in metal-catalyzed growth, enhancer effects on SWCNT growth from nonmetallic seeds have been scarcely studied, mainly due to the originally low efficiency of SWCNT growth particularly at high temperatures. We recently realized high-temperature SWCNT growth using ND-derived nanoparticles, which we hereafter call carbon nanoparticles (CNPs), as solid growth seeds [61]. The CNPs were prepared by transforming ND at high temperature, as will be described in detail in Method section. To efficiently obtain highly crystalline SWCNTs from the CNPs, we developed a two-step growth process [61]. In this process, the condition of the initial and secondary growth steps was separately adjusted to provide a suitable growth-driving force for SWCNT cap formation [62] and stationary nanotube elongation, respectively. We also reported a preliminary result on the etching of a-C by $H_2O$ in a CNP-based growth system [61]. A more thorough study on growth enhancers, focusing on a combination with different carbon sources, will be beneficial for understanding the mechanism of SWCNT growth from the unconventional growth seeds and realizing the efficient growth of structure-controlled and high-quality SWCNTs.

In this research, to understand the effects of enhancers on SWCNT growth from the CNPs, we systematically conducted SWCNT growth with the injection of $CO_2$ and $H_2O$ in $C_2H_2$- and ethylene ($C_2H_4$)-supplied SWCNT growth processes at high temperatures. The effects of $CO_2$ were first examined in a simple one-step growth process with the supply of $C_2H_4$ or $C_2H_2$ by changing the concentration and the supply delay of $CO_2$. The comparison of the results with the previous results on $H_2O$ revealed the similarities and differences in the roles of the two growth enhancers including the oxidation activity. Remarkably, $CO_2$ in combination with $C_2H_2$ enhanced a-C deposition as well as SWCNT growth, which revealed the occurrence of the dehydration reaction. We then applied $CO_2$ and $H_2O$ into the two-step growth process where the efficient growth of high quality SWCNTs was realized [61]. In addition to the concentration dependence of the enhancers, the change in the quality and yield of SWCNTs along the growth steps were investigated with different combinations of the growth enhancers and the carbon sources. We discuss the mechanism of the growth enhancers based on the change in chemical potential considering the dehydrogenation reaction between $C_2H_2$ and $CO_2$ as well as the oxidation reaction.

## 2. Experimental Section

### 2.1 Preparation of growth seeds

Figure 1 illustrates the entire process of SWCNT growth. As the starting material of growth seeds for SWCNTs, purified ND particles prepared via the detonation method [63] were dispersed in ethanol (2.0 wt%). The diameter of the ND particles was originally 10 nm, which was reduced to a smaller size in the following process. The impurity concentration of the ND particles was 2100, 80, and 6–20 ppm for Zr, Fe, and Pd, respectively. The SWCNTs were confirmed to grow from the ND-derived CNPs instead of the metal impurity contained in ND [61].

To support the growth seeds, ~10-mm$^2$ Si substrates with a 300-nm-thick thermal oxide layer were used. The substrates were cleaned via an ozone treatment process (L-UV253, Japan Electronics Industry) by flowing 6 l/min $O_2$ for 5 min under ultraviolet light for 60 min, followed by exhausting with 6 l/min $N_2$ for 5 min. After the cleaning of the Si substrates, 20 μl of the ND solution treated with ultrasonication was dropped on the surface of the substrates. Additionally, the same cleaned Si substrates without any growth seeds were used as a reference (hereafter called blank samples) to compare the amount of a-C deposited directly on them and on the SWCNT samples.

### 2.2 One-step growth process with $CO_2$ injection

To transform ND particles into the CNPs which are suitable for the growth seeds, pretreatment was conducted before the one-step and two-step SWCNT growth processes. The pretreatment of the growth seeds and the synthesis of SWCNTs were performed in a tubular CVD furnace (GE-1000, GII Techno) where a quartz tube chamber was used to load the substrates. The quartz tube chamber was designed to have a semicircular cross section with a diameter of 43.6 mm and heating-zone length of 890 mm. The substrates were placed at the center of the quartz tube chamber and treated with the surface-cleaning process in air for 10 min at 600 °C (Figure 1 (b), process 1). During this process, the diameter of ND was reduced to 2–3 nm, which is appropriate for SWCNT growth [61]. In the following annealing procedure, the heating temperature was gradually increased to 1000 °C and maintained for 1 h in Ar at 85 kPa with 20-sccm Ar injection. During this step, the surface of ND ($sp^3$ carbon) changed into more stable graphitic shells ($sp^2$ carbon), called carbon nano-onion structure [64, 65]. We used the CNPs possessing this structure as the growth seeds.

After the annealing process, SWCNTs were synthesized with two types of carbon feedstocks: $C_2H_2$ (2%)/Ar and $C_2H_4$ (2%)/Ar. Considering the difference in the thermal decomposition behaviors of $C_2H_2$ and $C_2H_4$ [66, 67], the synthesis condition was adjusted to fit the SWCNT growth from each carbon feedstock.

When using $C_2H_4$ as the carbon feedstock, SWCNTs were synthesized at 1000 °C for 10 min with a mixture of 2-sccm $C_2H_4$/Ar and 18-sccm $CO_2$/Ar while the total pressure was kept at 5 kPa, corresponding to a $C_2H_4$ partial pressure of 10 Pa. During the growth process, the partial pressure of $CO_2$ was varied from 0 to 180 Pa by changing the flow rates of $CO_2$ and Ar individually. When the partial pressure of $CO_2$ was 0 Pa, the carrier gas was $H_2$ (3%)/Ar instead of $CO_2$/Ar, for the usual condition for the SWCNT growth without a growth enhancer. In addition, in $C_2H_4$-supplied growth, $CO_2$ injection time was controlled. Figure 1 (a) shows that the start time (x min) for the injection of $CO_2$ at 140 Pa was varied as 0 (added from the beginning of the growth), 2, 5, 7, and 10 min (no $CO_2$ addition). Upon completion of the growth process, gas flows were switched to 20-sccm Ar and the temperature was decreased to room temperature while the total pressure was maintained at ~57 Pa.

When using $C_2H_2$ as the carbon feedstock, SWCNTs were synthesized at 900 °C for 10 min with a mixture of 1-sccm $C_2H_2$/Ar and 19-sccm $CO_2$/Ar. Since $C_2H_2$ is highly active, the total pressure was kept at 500 Pa, corresponding to a $C_2H_2$ partial pressure of 0.5 Pa, which is lower than the case of $C_2H_4$. During the growth process, the partial pressure of $CO_2$ contained in the 19-sccm $CO_2$/Ar was varied from 0 (0%) to 0.25 Pa (0.05%) by changing the flow rates of $CO_2$ and Ar individually. When the partial pressure of $CO_2$ was 0 Pa, the carrier gas was 19-sccm $H_2$ (3%)/Ar instead of 19-sccm $CO_2$/Ar.

## 2.3 Two-step growth process with $CO_2$ or $H_2O$ injection

As reported in the previous study [61], the SWCNT yield for growth at high temperatures is considerably increased by the two-step growth process. Therefore, we employed the two-step growth process in this study as well. As shown in Figure 1 (b), before the two-step growth, the pretreatment of the growth seeds is needed (process 1 for heating in air and process 2 for annealing in Ar). Then, the two-step growth process starts, including the initial growth step (process 3 for cap formation), the temperature-rising step (process 4), and the secondary growth step (process 5 for the stationary elongation). After the 1-h annealing process at 1000 °C (process 2 in Figure 1 (b)), $C_2H_2$ (2%)/Ar and $C_2H_4$ (2%)/Ar were used as the carbon feedstock, and the growth condition was adjusted for each growth step and feedstock. $H_2O$ or $CO_2$ was added as the growth enhancer after the initial growth stage (process 3 in Figure 1 (b)). Because the injection of $H_2O$ in $C_2H_2$-supplied growth has been studied in our previous work [61], the other three combinations of the feedstocks and enhancers were conducted in this study.

In the $C_2H_2$-supplied two-step growth process, the initial growth step was held at 850 °C for 1 min with a mixture of 2-sccm $C_2H_2$/Ar and 18-sccm $H_2$/Ar while the total pressure was maintained at 500 Pa, corresponding to a $C_2H_2$ partial pressure of 1 Pa. When the initial growth step was finished, the growth temperature was set to 1000 °C and the mass flow rate was adjusted to 1-sccm $C_2H_2$/Ar and 19-sccm $CO_2$/Ar, where the partial pressure of $C_2H_2$ was 0.5 Pa. The partial pressure of $CO_2$ added during the temperature-rising step (process 4 in Figure 1 (b)) was optimized to 0.05 Pa. Once the temperature reached 1000 °C, the secondary growth step began. The partial pressure of $CO_2$ was varied from 0.1 to 0.5 Pa by individually controlling the flow rates of $CO_2$ and Ar while the mass flow was maintained at 1-sccm $C_2H_2$/Ar and 19-sccm $CO_2$/Ar. When choosing $H_2O$ as the growth enhancer, which has been studied in the previous work [61], the partial pressure of injected $H_2O$ during the temperature-rising step was optimized to 0.15 Pa. Once the temperature reached 1000 °C, the secondary growth step commenced. The partial pressure of $H_2O$ was maintained at 0.15 Pa while the mass flow was maintained at 1-sccm $C_2H_4$/Ar and 19-sccm $H_2O$/Ar. For comparison, the supply of 19-sccm $CO_2$/Ar or $H_2O$/Ar was replaced with that of 19-sccm $H_2$ (3%)/Ar to realize SWCNT growth without growth enhancers.

In the $C_2H_4$-supplied two-step growth process, the initial growth step was held at 900 °C for 1 min with a mixture of 4-sccm $C_2H_4$/Ar and 16-sccm $H_2$/Ar while the total pressure was kept at 5 kPa, corresponding to a $C_2H_4$ partial pressure of 20 Pa. When the initial growth step was finished, the growth temperature was set to 1000 °C and the mass flow was adjusted to 1-sccm $C_2H_4$/Ar and 19-sccm $CO_2$/Ar, where the partial pressure of $C_2H_4$ was 5 Pa. Similar to the aforementioned $C_2H_2$-supplied case, the growth enhancer ($H_2O$ or $CO_2$) was injected when the initial growth step was finished. The partial pressure of $CO_2$ added during the temperature-rising step was optimized to 2 Pa. Once the temperature reached 1000 °C, the secondary growth step commenced. The partial pressure of $CO_2$ was varied from 2 to 62.5 Pa while the mass flow was maintained at 1-sccm $C_2H_4$/Ar and 19-sccm $CO_2$/Ar. When choosing $H_2O$ as the growth enhancer, the partial pressure of injected $H_2O$ during the temperature-rising step was optimized to 1 Pa. Once the temperature reached 1000 °C, the secondary growth step started. The partial pressure of $H_2O$ was varied from 0 to 1.5 Pa while the mass flow was maintained at 1-sccm $C_2H_4$/Ar and 19-sccm $H_2O$/Ar. For comparison, the injection of 19-sccm $CO_2$/Ar or $H_2O$/Ar was replaced with that of 19-sccm $H_2$ (3%)/Ar to realize SWCNT growth without growth enhancers.

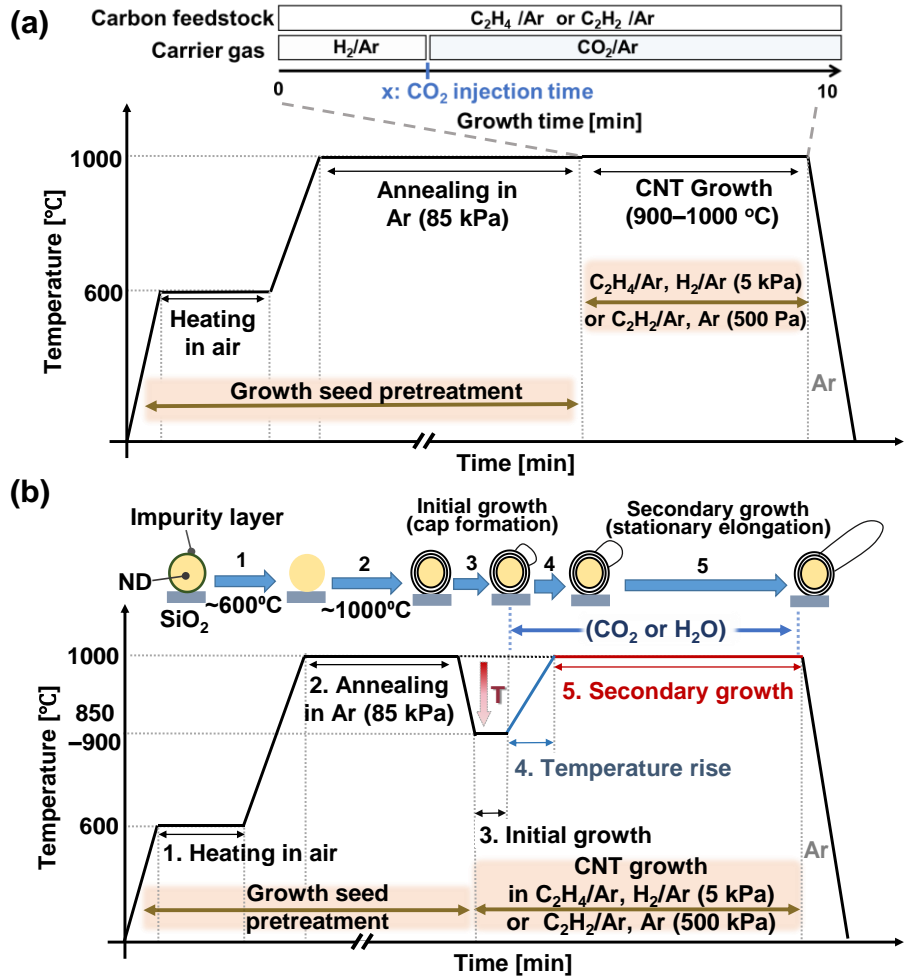

**Figure 1:** (a, b) Temperature profiles of the reaction furnace for (a) one-step growth process and (b) two-step growth process as a function of processing time. (a) The top part shows the gas composition profile during SWCNT growth. In the $C_2H_4$-supplied case, the variation of the $CO_2$ injection time (x = 0, 2, 5, 7, and 10 min) during the growth process was included. (b) The top part shows a schematic diagram of the two-step growth process of SWCNTs from the CNP growth seeds.

## 2.4 Structure characterization and yield evaluation of synthesized SWCNTs

We analyzed the structure of the synthesized SWCNTs using Raman spectroscopy. A Raman spectrometer (LabRAM HR800, HORIBA Jovin Yvon) was used with an excitation wavelength $\lambda_{ex}$ of 633 nm. The laser spot size was approximately 0.9 μm, and the laser power was approximately 7 mW at the measurement point. The exposure time of each measurement spot was 1 s for 5 cycles. Raman spectra of 30 randomly selected spots with SWCNT signal were collected from each sample, and the averaged spectra were used for the analysis. The quality of formed SWCNTs was evaluated using the intensity ratio of the G-band (~1590 cm$^{-1}$) to the D-band (1330–1360 cm$^{-1}$), which was represented as $I_G/I_D$ [68]. In most of the SWCNT growth situations, the density of SWCNTs was low in the high-temperature synthesis case and the G-band was occasionally not observed. Thus, the growth yield of SWCNTs was evaluated through the G-band observation frequency (GOF), that is, the number of appearances of the G-band in the measured spectra divided by the total number of measurements. Notably, in some growth cases, the density of SWCNTs was high and the GOF was approximately 100%, so that the difference of the growth yield could not be reflected by the changes in the GOF. In such a situation, the SWCNT quantity was evaluated by comparing the intensity ratio of the G-band (~1590 cm$^{-1}$) with the Raman peak from Si substrates (~520 cm$^{-1}$), which was represented as $I_G/I_{Si}$. Typically, Raman spectra of 100 randomly selected spots (with or without SWCNT signal) were collected for the evaluation of the GOF. More measurement spots were used to determine the GOF than that for averaging Raman spectra to ensure the accuracy of the GOF even for samples with low-density SWCNTs. Scanning electron microscopy (SEM) (NVision, Carl Zeiss and S-4800, Hitachi) was used for morphology observations of SWCNTs with an acceleration voltage of 1−6 kV.

## 3. Results and Discussion

### 3.1. Effects of $CO_2$ on $C_2H_4$-supplied one-step growth

To investigate the effect of growth enhancers on SWCNT growth without metal catalysts, $CO_2$ was injected in the high-temperature synthesis process with the CNPs as the growth seeds and different carbon feedstocks, namely, $C_2H_2$ and $C_2H_4$. Figure 2 depicts the variation of SWCNT yield and quality for various concentrations of $CO_2$ input into the $C_2H_4$-supplied growth processes.

By injecting $CO_2$ from the beginning of the growth, the SWCNTs were synthesized from $C_2H_4$ at 10 Pa and 1000 °C for 10 min. The Raman spectra shown in Figure 2(a), which are normalized to the Si peak intensity at ~520 cm$^{-1}$, show the variation of the D-band and G-band with the injection of $CO_2$ at different partial pressures (0–180 Pa). The quality ($I_G/I_D$ ratio) and quantity (GOF) of SWCNTs are shown in Figure 2 (b). In addition, the $I_D/I_{Si}$ ratio of the SWCNT and blank samples shown in Figure 2 (b) helps to distinguish the origin of the D-band: a-C deposited on samples or defects in SWCNT lattices. In the blank samples, since no SWCNTs are grown, the $I_D/I_{Si}$ ratio reflects only the degree of a-C formation under the growth condition. If a-C is deposited on a blank sample, a similar or higher density of a-C should be deposited on an SWCNT sample under the same condition due to the curvature of nanotubes and the increased surface area of the sample. If a-C is not deposited on a blank sample, a-C deposition on an SWCNT sample should be very low or negligible under the same condition. Upon the addition of $CO_2$, the grown SWCNTs exhibit a decrease in D-band intensity and a slight change in G-band intensity (Figure 2 (a)). An improvement in the SWCNT quality is apparently revealed by the increase in the $I_G/I_D$ ratio (Figure 2(b)). The decreasing tendency observed in the $I_D/I_{Si}$ ratio from the blank samples represents the decrease in a-C. A similar tendency is exhibited by the $I_D/I_{Si}$ ratio from the SWCNT samples, and finally the $I_D/I_{Si}$ ratio is close to that of the blank samples. These results prove that the origin of the D-band from the SWCNTs should be mostly from a-C, and with the injection of $CO_2$, a-C deposition is gradually prevented. The averaged $I_G/I_D$ ratio reached approximately 100 when the partial pressure of $CO_2$ was adjusted to 160 Pa. With further increase in the partial pressure of $CO_2$ to an excess value, 180 Pa, the $I_D/I_{Si}$ ratio did not increase in the blank samples but did slightly increase in the SWCNT samples. Thus, the increase in the $I_D/I_{Si}$ ratio in the SWCNT samples reflects the formation of defects. Meanwhile, the yield of SWCNTs shown as the GOF in Figure 2 (b) exhibited a small reduction with the injection of $CO_2$, until an excess amount of $CO_2$ (180 Pa) was added. The variation in the growth yield indicates that the injection of $CO_2$ also influences the growth efficiency of SWCNTs. This growth result indicates the potential of $CO_2$ in preventing the deposition of a-C at high temperatures. Additionally, the slight formation of defects even with excess $CO_2$ confirms the mild etching ability of $CO_2$. It should be noted that, besides Raman spectroscopy, transmission electron microscopy (TEM) is another promising method to analyze the structure of SWCNTs and a-C. However, it is still challenging to prepare

SWCNT samples on TEM grids because of the relatively low yield of SWCNTs in this study.

Next, the etching effect of $CO_2$ on SWCNT growth was analyzed by delaying the injection time of $CO_2$ to 2, 5, and 7 min while the partial pressure was maintained at 140 Pa. With the delaying of the injection time, the D-band intensity gradually increased, as shown in Figure 2 (c). A decrease in the $I_G/I_D$ ratio can be seen in Figure 2 (d), which reflects the reduction in the SWCNT quality. Moreover, the increase in the $I_D/I_{Si}$ ratio of both SWCNTs and the corresponding blank samples, as seen in Figure 2 (d), proves that the amount of a-C deposited on the SWCNT samples gradually increases. Even with an increase in the partial pressure of $CO_2$ to 200 Pa, when delaying the injection time to 5 min, the increase in a-C deposition cannot be prevented perfectly (shown in Figure S1). The milder a-C-etching ability of $CO_2$ has been discussed in previous studies on metal-catalyzed growth systems [51-57, 59, 60]. Moreover, without a metal catalyst, the increase in the D-band intensity along with the delaying of the injection time reflects that the a-C etching rate of $CO_2$ is limited to be of the same order as the deposition rate of a-C or even slower. Notably, the GOF in Figure 2 (d) shows a slightly decreased yield when $CO_2$ is added from the beginning or after 2 min of growth. Because the initial cap formation is expected to occur in the first few minutes during the synthesis, such a slight decrease in yield implies that $CO_2$ reduces the growth yield of SWCNTs by influencing the initial growth efficiency. Associating the decrease in the growth yield with the a-C-etching ability, we assume that during SWCNT growth from the CNP growth seeds, $CO_2$ reduces the density of carbon species adsorbed on the surface of growth seeds via an oxidation reaction. Such a decrease in the density of activated carbon species adsorbed on the growth seeds reduced the carbon supply rate to the growth edge, which explains both the reduction in the growth yield and the deposition of a-C. Thus, in this $CO_2$-participated SWCNT growth process, the limited deposition of a-C is due to the reduction of the activated carbon species instead of the etching of a-C. Unavoidably, such a decrease in carbon supply also slightly influences the SWCNT growth efficiency, particularly in the nucleation stage, which is more sensitive to the growth condition than is the stationary elongation stage.

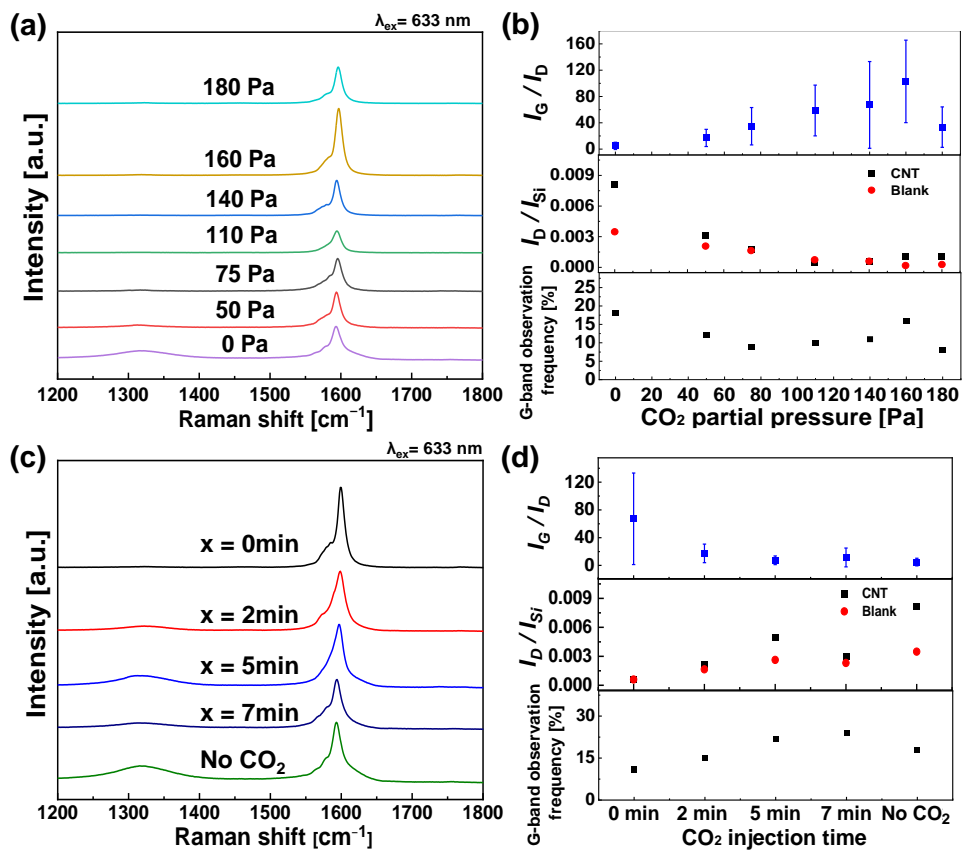

**Figure 2:** (a) Raman spectra of SWCNTs obtained by one-step growth from $C_2H_4$ with different partial pressures of $CO_2$ injected from the commencement of growth. (b) GOF, $I_D/I_{Si}$ ratio, and $I_G/I_D$ ratio of the same samples as (a). The error bars presented in the $I_G/I_D$ plot represent the standard deviation of the value from different measurement spots. $I_D/I_{Si}$ was plotted for the blank samples as well. (c) Raman spectra of SWCNTs grown with different delay of $CO_2$ supply at 140 Pa: 0, 2, 5, and 7 min (indicated as x), and without $CO_2$ addition (No $CO_2$). (d) Corresponding GOF, $I_D/I_{Si}$, and $I_G/I_D$. Standard deviation was also calculated and presented as the error bars of averaged $I_G/I_D$.

### 3.2. Effects of $CO_2$ on $C_2H_2$-supplied one-step growth

We investigated $CO_2$-assisted SWCNT growth using $C_2H_2$ at 0.5 Pa as the carbon source at 900 °C for 10 min. $CO_2$ was added from the beginning of SWCNT growth. As shown in Figure 3 (a), the Raman spectra exhibited the difference in the D-band intensity when the partial pressure of $CO_2$ was changed from 0.025 to 0.25 Pa. In contrast to the D-band suppression in the $C_2H_4$-supplied case (Figure 2), a decrease in the $I_D/I_{Si}$ ratios of the SWCNT samples was not clearly observed, and the $I_D/I_{Si}$ ratios of the blank samples were not negligible even with the highest concentration of $CO_2$ in this case (Figure 3 (b)). This result indicates that the a-C etching by $CO_2$ in the $C_2H_2$-supplied SWCNT growth was not evident. A relatively minor improvement in SWCNT quality can be observed in the $I_G/I_D$ shown in Figure 3 (b), which is mainly caused by the suspension of a-C deposition.

Compared with less active $C_2H_4$, the use of highly active $C_2H_2$ as the carbon source at high growth temperatures resulted in a higher growth yield. The GOF of most samples reached 100% except for the sample with the highest $CO_2$ pressure under this experiment condition, as shown in Figure 3 (a). Thus, $I_G/I_{Si}$ was used to evaluate the yield in this case, which is shown in Figure 3 (b). In contrast to the case of $C_2H_4$, the increased yield was achieved with $C_2H_2$ when the partial pressure of $CO_2$ was set to 0.05 and 0.15 Pa. Because a similar $I_G/I_D$ ratio was obtained with and without the assistance of $CO_2$, the increase in growth yield due to $CO_2$ is unlikely to be caused by the prevention of a-C deposition, which could prolong the lifetime of the growth seeds. Thus, we need to consider a possible role of $CO_2$ other than the etching of a-C.

In $C_2H_2$-supplied CNT growth, besides etching ability, $CO_2$ has been recently proven to exhibit a promoting role in the dehydrogenation reaction of hydrocarbon, which finally increases CNT yield [53]. In the $CO_2$-enhanced CNT growth process, two chemical reactions are possible between $C_2H_2$ and $CO_2$:

$$C_2H_2 + CO_2 \leftrightarrow 2C + H_2O + CO, \qquad (1)$$

$$C_2H_2 + CO_2 \leftrightarrow C + H_2 + 2CO. \qquad (2)$$

Based on these reactions, a reduced activation energy barrier (from 5.58 down to 4.97 eV) of the first H dissociation in the $C_2H_2$ dehydrogenation process with the injection of $CO_2$ was illustrated through a simulation method [59]. Similar to these studies, we assume that the dehydrogenation reactions between $C_2H_2$ and $CO_2$ also occur in the SWCNT growth process from the nonmetallic growth seeds. Based on the enhanced dehydrogenation reaction, the formation of activated carbon species on the growth seeds was promoted and caused a higher carbon-source supply, which resulted in an increase in both SWCNT growth yield and a-C deposition.

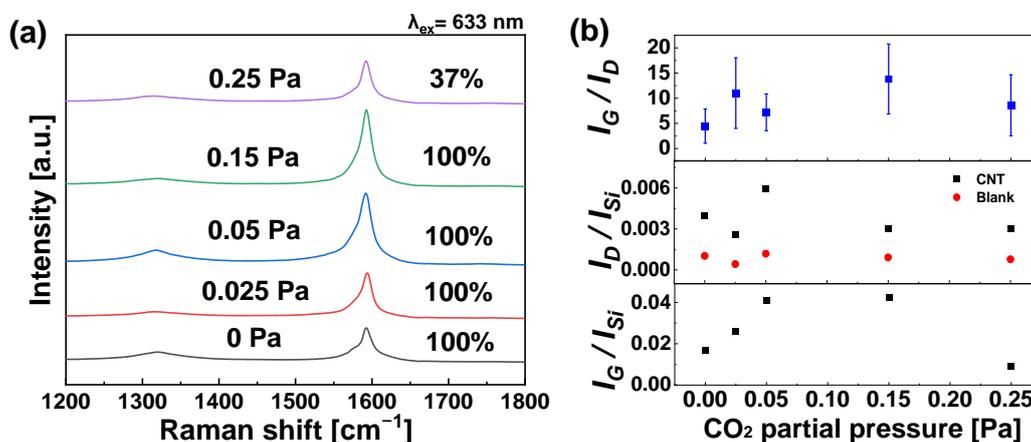

**Figure 3:** (a) Raman spectra of SWCNTs obtained by one-step growth from $C_2H_2$ with different partial pressures of $CO_2$ injected from the commencement of growth. The GOF percentage is indicated in the graph. (b) $I_G/I_{Si}$, $I_D/I_{Si}$, and $I_G/I_D$ ratios of the same samples as (a). The error bars represent the standard deviation.

### 3.3. Effects of enhancers on $C_2H_4$-supplied two-step growth

Based on the enhancing effect of $CO_2$ in one-step SWCNT synthesis from the CNP growth seeds, we applied $CO_2$ to the two-step growth process, which was proven to increase the yield of highly crystalline SWCNTs [61]. In the two-step growth system shown in Figure 4 (a), the $C_2H_4$-supplied growth SWCNTs using $H_2O$ as an enhancer were also conducted for comparison.

Figures 4 (b) and (c) present the quality and quantity variation of SWCNTs from the $C_2H_4$-supplied two-step growth process when using different growth enhancers, namely, $H_2O$ and $CO_2$. Regarding the injection time of the growth enhancers, we previously found that additional defects were formed if $H_2O$ was injected from the initial growth step [61]. Even when using $CO_2$ as the growth enhancer, the supply of $CO_2$ from the initial growth step resulted in a slight decrease in SWCNT yield in the one-step growth from $C_2H_4$ (Figure 2 (c) and (d)). Thus, to maintain the nucleation efficiency during the initial growth step (process 1 in Figure 4 (a)), both enhancers were added at the beginning of the temperature-rising step (process 2 in Figure 4 (a)). The partial pressure of the growth enhancers during the temperature-rising step was optimized (1 Pa for $H_2O$ or 2 Pa for $CO_2$) to prevent the deposition of a-C (Figure S3). The influence of enhancers was analyzed according to the changes in the yield and quality of SWCNTs when varying the partial pressure of $H_2O$ and $CO_2$ at the secondary growth step (process 3 in Figure 4 (a)). The condition is represented by the partial pressure of enhancers during the three processes; for example, 0-1-1 Pa means that the partial pressure during the initial growth step, temperature-rising step, and secondary growth step is 0, 1, and 1 Pa, respectively.

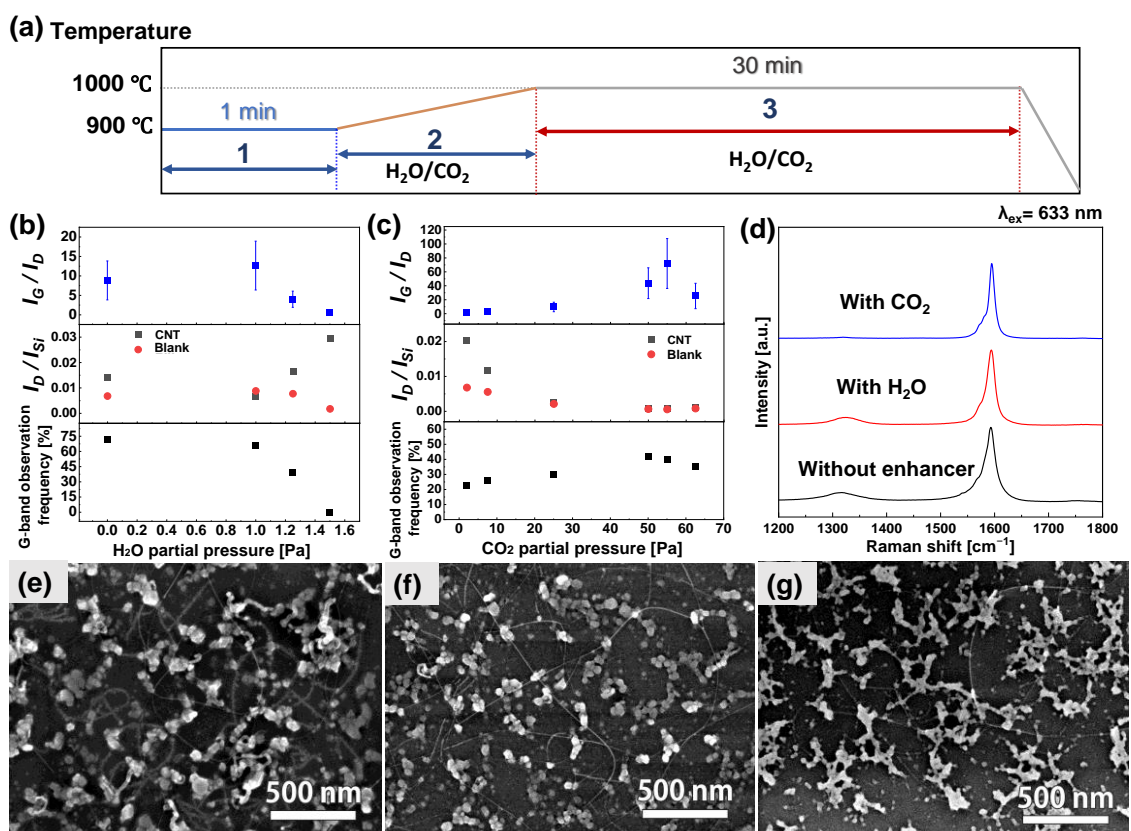

**Figure 4:** (a) Temperature profiles for the $C_2H_4$-supplied SWCNT two-step growth process with growth enhancers as $H_2O$ or $CO_2$. Processes 1, 2, and 3 indicated in the profile represent the initial growth step, temperature-rising step, and secondary growth step, respectively. (b, c) GOF, $I_D/I_{Si}$ ratio, and $I_G/I_D$ ratio of the SWCNTs grown from $C_2H_4$ with different partial pressures of (b) $H_2O$ and (c) $CO_2$ during the secondary growth step. The partial pressures of $H_2O$ and $CO_2$ in the secondary growth step were varied from 0 to 1.5 Pa and 2 to 62.5 Pa, respectively. Further, the partial pressures during the temperature-rising step were fixed to 1 and 2 Pa for $H_2O$ and $CO_2$, respectively. Standard deviation was also calculated and presented as the error bar of averaged $I_G/I_D$. (d) Raman spectra and (e, f, g) SEM images (acceleration voltage: 5 kV) of SWCNTs synthesized (e) without a growth enhancer, (f) with the injection of $H_2O$ (0-1-1 Pa), and (g) with the injection of $CO_2$ (0-2-55 Pa).

In Figure 4 (b), with increasing $H_2O$ partial pressure at the secondary growth step from 0 to 1.5 Pa (0-1-0, 0-1-1, 0-1-1.25, and 0-1-1.5 Pa), the averaged $I_G/I_D$ ratios exhibit a decreasing tendency, which indicates an increase in a-C or defect formation caused by excess $H_2O$. To elucidate the origin of the D-band, we compared the averaged $I_D/I_{Si}$ ratios of the SWCNT samples and the corresponding blank samples. For the growth process in which no $H_2O$ or only 1-Pa $H_2O$ (0-1-0 Pa and 0-1-1 Pa) was supplied during the secondary growth step, slightly higher or similar $I_D/I_{Si}$ ratios were observed for the SWCNT samples compared with the blank samples. The appearance of the D-band in the blank samples originated from the deposition of a-C. Thus, below 1 Pa of $H_2O$, the appearance of the D-band from SWCNT samples is explained by the deposition of a-C. When the partial pressure of $H_2O$ was increased to 1.5 Pa at the secondary growth step (0-1-1.5 Pa), $I_D/I_{Si}$ decreased in the blank samples, which indicates the removal of a-C. However, a dramatic increase in $I_D/I_{Si}$ was found in the SWCNTs samples. Such a D-band of the SWCNTs samples should originate from the formation of defects. Therefore, we assume that when the partial pressure of $H_2O$ was greater than 1 Pa in this growth system, in addition to a-C, SWCNT structures were partially etched by $H_2O$, resulting in defect formation. In addition, when the partial pressure of $H_2O$ was increased to 1.25 Pa, the GOF decreased to ~40%. The GOF further decreased to 0% when the partial pressure of $H_2O$ increased to 1.5 Pa. The increase in D-band intensity represents the formation of the $sp^3$ structure (defects), and the disappearance of the G-band represents the loss of SWCNTs. This dramatic decrease in SWCNT yield also reflects one of the disadvantages of $H_2O$, which comes from its strong oxidizing ability at high temperatures. Because of this strong oxidizing ability, the oxidizing reaction occurred not only in a-C but also in SWCNTs and activated carbon species, particularly with excess $H_2O$, which caused a decrease in the SWCNT quality and quantity. The abovementioned growth results indicate that the complete removal of a-C is difficult by adding $H_2O$ without causing damage to SWCNTs and the partial pressure of $H_2O$ needs to be accurately controlled to avoid the destruction of SWCNTs.

Moreover, using $C_2H_4$ as the carbon source, a $CO_2$-assisted two-step growth was conducted; the corresponding results are summarized in Figure 4 (c). Because of its milder oxidizing ability, $CO_2$ with a partial pressure greater than that of $H_2O$ was used for the secondary growth step. With the same growth condition (carbon-source partial pressure, growth time, and growth temperature) as that using $H_2O$, $CO_2$ exhibits a higher effectivity in improving the quality of SWCNTs according to the increase in the averaged $I_G/I_D$ ratios shown in Figure 4 (c). Additionally, with an increase in the $CO_2$ partial pressure from 2 to 62.5 Pa (0-2-2, 0-2-7.5, 0-2-25, c, 0-2-55, and 0-2-62.5 Pa), a decreasing tendency of the averaged $I_D/I_{Si}$ ratio was found in both the SWCNT samples and the blank samples. Such an increase in the $I_G/I_D$ ratio to ~69 and a decrease of $I_D/I_{Si}$ ratio close to 0 indicates the formation of highly crystalline SWCNTs with negligible deposition of a-C. The lower deposition of a-C helps to keep the surface of the growth seeds clean and prolongs the growth lifetime, which explains the slight increase in the growth yield. Figure 4 (d) depicts the Raman spectra of the SWCNTs grown under optimized growth conditions without growth enhancers, with $H_2O$ (0-1-1 Pa), and with $CO_2$ (0-2-55 Pa). Because the G-band intensity in the Raman spectra is normalized, the difference in the D-band intensity can be clearly observed. Compared with the growth without growth enhancers or with $H_2O$, $CO_2$ presents its advantage in preventing a-C deposition when using $C_2H_4$ as the carbon source. Figures 4 (e), (f), and (g) show the SEM images of the SWCNTs grown without growth enhancers, with $H_2O$, and with $CO_2$, respectively. The particles shown in the SEM images are agglomerates of the CNPs. In Figure 4 (e), without growth enhancers, SWCNTs with a significantly shorter length (~50 nm) appeared owing to the deposition of a-C, which terminated the growth in a short time. When $H_2O$ is added to the growth process, the decrease in the short SWCNTs shown in Figure 4 (f) represents the suppression of a-C deposition, which prolongs the lifetime of the growth seeds. When using $CO_2$ as the growth enhancer, the formation of a-C was prevented more effectively; longer SWCNTs can be frequently found in Figure 4 (g).

We then investigated the changes in growth yield and quality along the process flow of the two-step growth with and without the growth enhancers (0-1-1 Pa of $H_2O$ or 0-2-55 Pa of $CO_2$). Raman spectra were acquired from the samples immediately after the three steps: the initial growth step, temperature-rising step, and secondary growth step. The changes in GOF and $I_G/I_D$ are shown in Figure 5. Even without growth enhancers, the slight increase in the $I_G/I_D$ ratios from the initial growth step to the temperature-rising step proves that the deposition of a-C is still low before the growth temperature rises to 1000 °C. The addition of $H_2O$ or $CO_2$ with the optimized conditions during the temperature-rising step causes an increase in the $I_G/I_D$ ratios compared with that without growth enhancers. This means that the deposition of a-C is further prevented by $H_2O$ or $CO_2$. However, according to Figure 5 (a), compared to the sample without growth enhancers, the decreased yield of SWCNTs was observed when $H_2O$ or $CO_2$ was added. Particularly for $CO_2$, there was almost no increase in yield during the temperature-rising step. Based on the role of $CO_2$ at high temperatures described in section 3.1 (one-step growth), we assume that the prevention of a-C formation is mainly caused by the decrease in carbon species adsorbed on the growth seeds. Such a decrease also caused the decrease in SWCNT yield.

After the temperature-rising step, the secondary growth step was conducted, in which the stationary growth of SWCNTs occurs. Notably, with the temperature rising from 900 °C to 1000 °C, the carbon-source decomposition rate increases, which causes an increase in the supply of activated carbon species [66]. Without the injection of growth enhancers, the deposition of a-C was enhanced, as shown in Figure 5 (b). The use of $H_2O$ could not provide SWCNTs with high crystallinity and low a-C simultaneously. The a-C deposition caused the decrease in the $I_G/I_D$ ratio and finally terminated the growth at the beginning of the secondary growth step, which is evidenced by the negligible increase in the GOF from the temperature-rising step to the secondary growth step in Figure 5 (a). By contrast, in the case of $CO_2$ addition, the partial pressure of $CO_2$ could be largely increased along with the increase in the supplied carbon source without damaging the SWCNTs because of its mild oxidizing ability. The increase in the GOF from the temperature-rising step to the secondary growth step indicates the enhancement effect of $CO_2$. Based on such enhancement results, $CO_2$ was further applied to the two-step growth with the supply of higher-partial-pressure $C_2H_4$ at the secondary growth step. Through the adjustment of the growth temperature at the secondary growth step, the $I_G/I_{Si}$ ratio was increased to ~0.15 while the GOF was 100%, and the $I_G/I_D$ ratio was kept at ~57 (Figure S4). We assume that $CO_2$ prolongs the lifetime of the growth seeds by keeping their surface clean and maintains the efficient growth of SWCNTs during the secondary growth step.

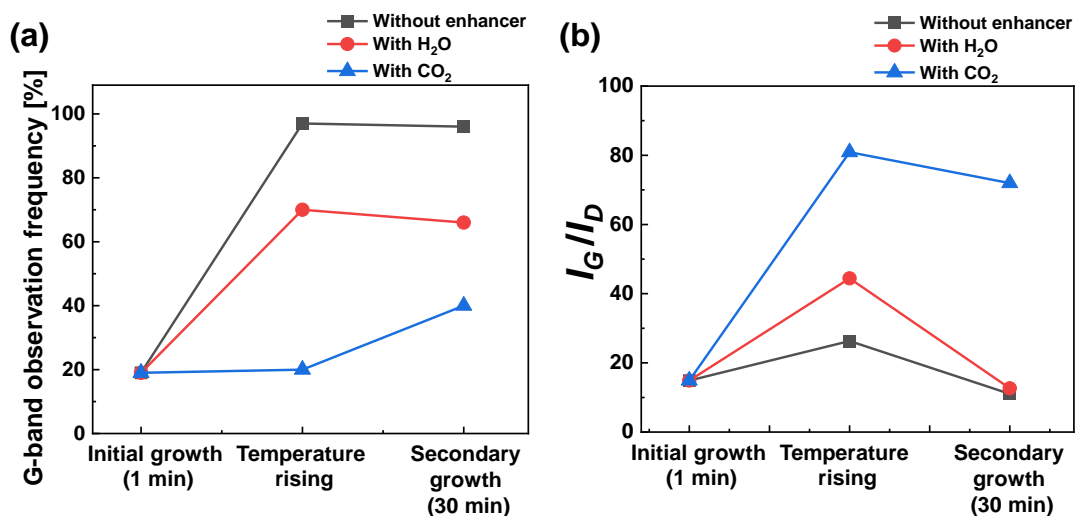

**Figure 5:** Changes in (a) yield (GOF) and (b) quality ($I_G/I_D$ ratio) of SWCNTs along the growth steps from $C_2H_4$ without growth enhancers, with $H_2O$ (0-1-1 Pa), and with $CO_2$ (0-2-55 Pa).

### 3.4. Effects of enhancers on $C_2H_2$-supplied two-step growth

We investigated the effect of $CO_2$ on the $C_2H_2$-supplied two-step growth process. The growth process is schematically shown in Figure 6 (a). Similar to the condition in the $C_2H_4$-supplied growth described in section 3.3, $CO_2$ was injected after the first growth step and its partial pressure during the temperature-rising step was optimized to 0.05 Pa. Because the growth result with the injection of $H_2O$ has already been discussed in our previous work [61], the growth results with $CO_2$ injection are mainly discussed in this section.

Figure 6 (b) depicts the variation of the quality ($I_G/I_D$ and $I_D/I_{Si}$ ratios) and yield (GOF) of SWCNTs with a change in the partial pressure of $CO_2$ at the secondary growth step from 0.1 to 0.5 Pa (0-0.05-0.1, 0-0.05-0.25, 0-0.05-0.35, and 0-0.05-0.5 Pa). The relevant Raman spectra are shown in Figure S5. Interestingly, contrary to the expected role of $CO_2$ for a-C deposition prevention, relatively low $I_G/I_D$ ratio were observed. Within the range of $CO_2$ partial pressure, the growth yield decreased with increasing $CO_2$. This indicates that the injection of $CO_2$ cannot prevent the deposition of a-C in the $C_2H_2$-supplied growth process. Even after the partial pressure of $CO_2$ was increased to 0.5 Pa (0-0.05-0.5 Pa), the yield decreased while the $I_G/I_D$ ratio did not change significantly, which means that the intensities of both the G-band and D-band decreased. However, as shown in Figure 6 (c), the D-band and G-band intensities of the sample with 0.1-Pa $CO_2$ (0-0.05-0.1 Pa) are higher than those of the sample without growth enhancers. Because the GOF of both samples is already 100% under this condition, the intensity of the G-band reflects the SWCNT yield. Moreover, according to the SEM images shown in Figures 6 (d) and (e), a higher density of SWCNTs was observed in the $CO_2$-injected sample than in the sample without growth enhancers. To analyze the increase of D-band intensity, we compared the change in a-C deposition on the blank samples along the two-step growth process with and without the injection of $CO_2$. As shown in Figure S6, higher $I_D/I_{Si}$ ratio was found in the $CO_2$-injected case. The increase of $I_D/I_{Si}$ should be attributed to the a-C deposition since there is no possibility of defect formation on the blank samples, where no SWCNTs were grown. Thus, for the sample grown with a small amount of $CO_2$, the higher G-band intensity represents a higher yield and the higher D-band intensity could reflect the more intense deposition of a-C. Moreover, the appearance of the radial breathing mode peak at ~180 cm$^{-1}$, which was not observed in the samples grown without a growth enhancer, indicates that thinner SWCNTs were formed with the injection of $CO_2$. The formation of the thinner SWCNTs should be due to the increased growth-driving force [30]. Considering that the growth temperature and partial pressure of $C_2H_2$ were fixed, the increased density of activated carbon species is regarded as the main factor that causes the increase in the growth-driving force. The activated carbon species are formed from the decomposition of $C_2H_2$. As discussed in section 3.2, we assume that $CO_2$ in $C_2H_2$-supplied growth mainly works as a growth enhancer, which promotes the dehydrogenation reaction of the carbon source and increases the density of activated carbon species. Such an increase in activated carbon species promotes SWCNT growth but also leads to uncontrolled a-C deposition.

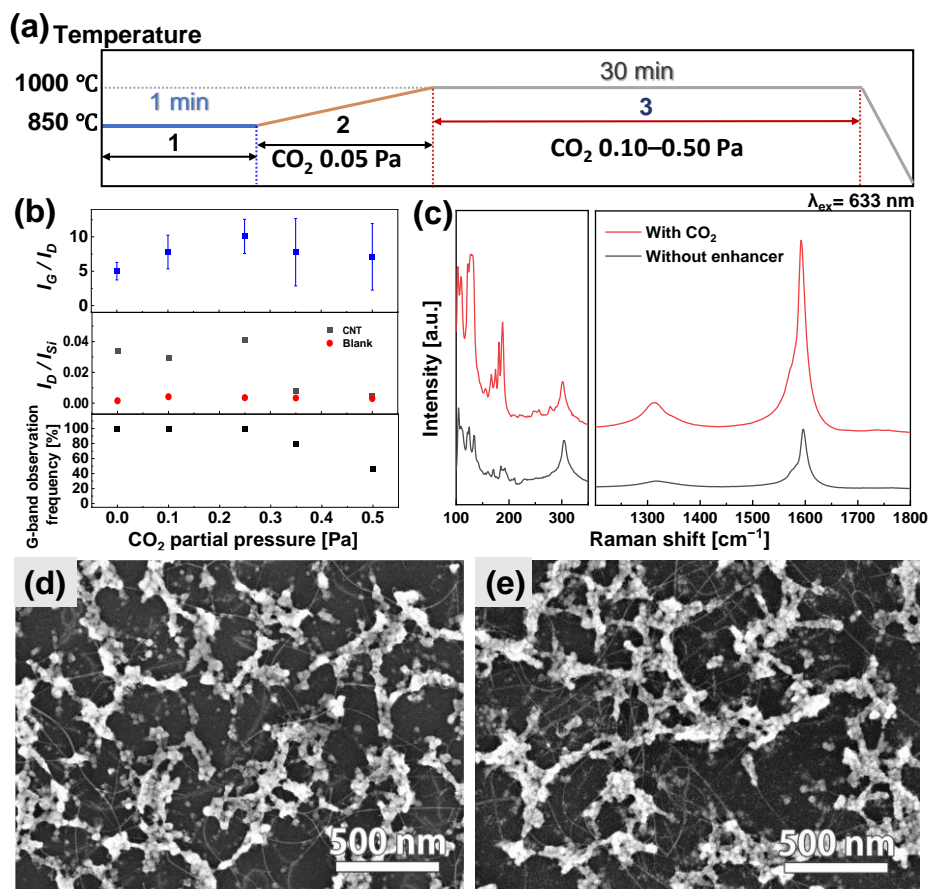

**Figure 6:** (a) Temperature profiles for the $C_2H_2$-supplied SWCNT two-step growth process with the injection of $CO_2$. (b) GOF, $I_D/I_{Si}$ ratio, and $I_G/I_D$ ratio of the SWCNTs grown from $C_2H_2$ with different partial pressures of $CO_2$ injected since the temperature-rising step. The partial pressure of $CO_2$ in the secondary growth step was varied from 0 to 0.5 Pa. The partial pressure of $CO_2$ in the temperature-rising step was fixed to 0.05 Pa. Standard deviation was also calculated and presented as the error bar of averaged $I_G/I_D$. (c) Raman spectra of the SWCNT samples synthesized with and without the injection of $CO_2$. The partial pressure of $CO_2$ was 0.05 and 0.1 Pa during the temperature-rising step and the secondary growth step, respectively (0-0.05-0.1 Pa). (d, e) SEM images (acceleration voltage: 5 kV) of the SWCNT samples synthesized (d) without the injection of a growth enhancer and (e) with the injection of $CO_2$ (0-0.05-0.1 Pa).

The changes in the quality and yield of SWCNTs along the growth steps with and without the injection of enhancers are shown in Figure 7. The result with $H_2O$ (0-0.15-0.15 Pa) has been taken from our previous work [61]. Raman measurement was conducted after the initial growth, temperature-rising, and secondary growth steps as in the case of $C_2H_4$. Because of the higher activity of $C_2H_2$ than $C_2H_4$, a-C was easily formed during the temperature-rising and secondary growth steps, which caused the decrease in the $I_G/I_D$ ratios when no growth enhancer was injected. As reported in our previous work [61], with accurate control of the $H_2O$ partial pressure (0-0.15-0.15 Pa), a-C was effectively removed and the $I_G/I_D$ ratio increased. At the same time, a part of the activated carbon species was also removed from the growth seed surface by $H_2O$, which resulted in the decrease in the SWCNT yield compared with the sample without growth enhancers, as shown in Figure 7 (a). When using $CO_2$ as the growth enhancer, contrasting results were obtained. The $I_G/I_D$ ratio decreased while the GOF was increased by $CO_2$. The dramatic increase in the GOF also supports our assumption that the carbon-source supply is promoted by the injection of $CO_2$. This promotion effect causes the increase in SWCNT yield as well as the deposition of a-C. Thus, based on the discussion above, we conclude that $CO_2$ plays more than one role in SWCNT growth from the CNP growth seeds depending on the employed carbon feedstocks.

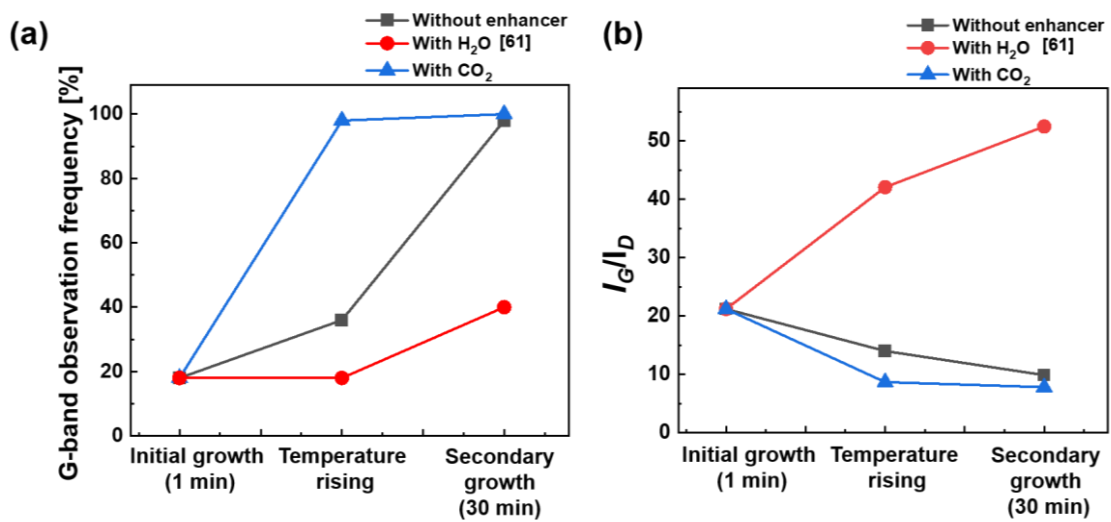

**Figure 7:** Changes in (a) yield (GOF) and (b) quality ($I_G/I_D$ ratio) of SWCNTs along the growth steps from $C_2H_2$ without growth enhancers, with $H_2O$ (0-0.15-0.15 Pa) [61], and with $CO_2$ (0-0.05-0.1 Pa).

## 3.5. Mechanism of growth enhancers in combination with carbon sources

The mechanism of the effect of enhancers ($H_2O$ and $CO_2$) on SWCNT growth behavior is derived from the observed results and is schematically depicted in Figure 8. In the SWCNT growth system based on the CNP growth seeds, carbon-source molecules in the gas phase first overcome the energy barrier to proceed with the dehydrogenation reaction, and activated carbon species are adsorbed on the surface of growth seeds. Then, the activated carbon species are attached to the edge of the SWCNT sidewall. Therefore, the conversion process of carbon-source molecules to SWCNTs is divided into two parts: the process in which the carbon-source molecules in the gas phase are transformed into activated carbon species of the adsorbed phase and the subsequent process in which the activated carbon species are transformed into the SWCNT structure. The activation energies ($E_A$) for these two processes are denoted as $E_{A1}$ and $E_{A2}$, respectively.

We first discuss the effect of $H_2O$. In SWCNT growth from both $C_2H_2$ and $C_2H_4$ using the CNP growth seeds, $H_2O$ was found to mainly prevent the formation of a-C. We assume that the preventing role is realized through the oxidizing reaction, which is similar to its role in SWCNT growth from a metallic catalyst. With the addition of $H_2O$, the chemical potential of activated carbon species decreased, which caused the decrease of SWCNT growth efficiency, as shown in Figures 8 (a) and (b). In addition to a-C, a part of the activated carbon species on the surface of growth seeds is etched away by $H_2O$, which decreases the density of the carbon species and accordingly their chemical potential. The decrease in the chemical potential increases the activation energy needed for the formation of SWCNTs ($E_{A2}$' in Figures 8 (a) and (b)), assuming that the transition state is preserved by the addition of $H_2O$. This change reasonably explains the decrease in the SWCNT growth efficiency in both the $C_2H_2$- and $C_2H_4$-supplied system by $H_2O$.

When using $CO_2$ as the growth enhancer, the reaction between the carbon source and $CO_2$ needs to be considered depending on the carbon source. In the case of $C_2H_4$, as the reaction with $CO_2$ is not significant, the main role of $CO_2$ is etching. With the injection of $CO_2$, the chemical potential of activated carbon species decreased, which caused the decrease of SWCNT growth efficiency, as shown in Figures 8 (a). Because the oxidizing ability of $CO_2$ is lower than that of $H_2O$, $CO_2$ mainly etches away activated carbon species on the growth seeds and does not directly react with the formed a-C. The decrease in the density of the carbon species causes a decrease in chemical potential and increase in activation energy, resulting in the low growth efficiency of SWCNTs. In the case of $C_2H_2$, a dehydrogenation reaction occurs between $CO_2$ and $C_2H_2$. As shown in Figure 8(b), the activation energy of the dehydrogenation reaction of $C_2H_2$ decreases from $E_{A1}$ (the process without $CO_2$) to $E_{A1}'$ (the process with $CO_2$). Such a decrease in the energy barrier helps to increase the rate of carbon feedstock transformation from the gas phase to the adsorbed phase. The increased adsorbed carbon specie density causes the increase in chemical potential of carbon and then the decrease in the activation energy for SWCNT formation ($E_{A2}''$). This growth condition further promotes the formation of both SWCNTs and a-C. Therefore, the reaction between $C_2H_2$ and $CO_2$ significantly affects the enhancement of SWCNT growth not only in the metal-catalyzed SWCNT growth but also in the solid carbon seed-based growth.

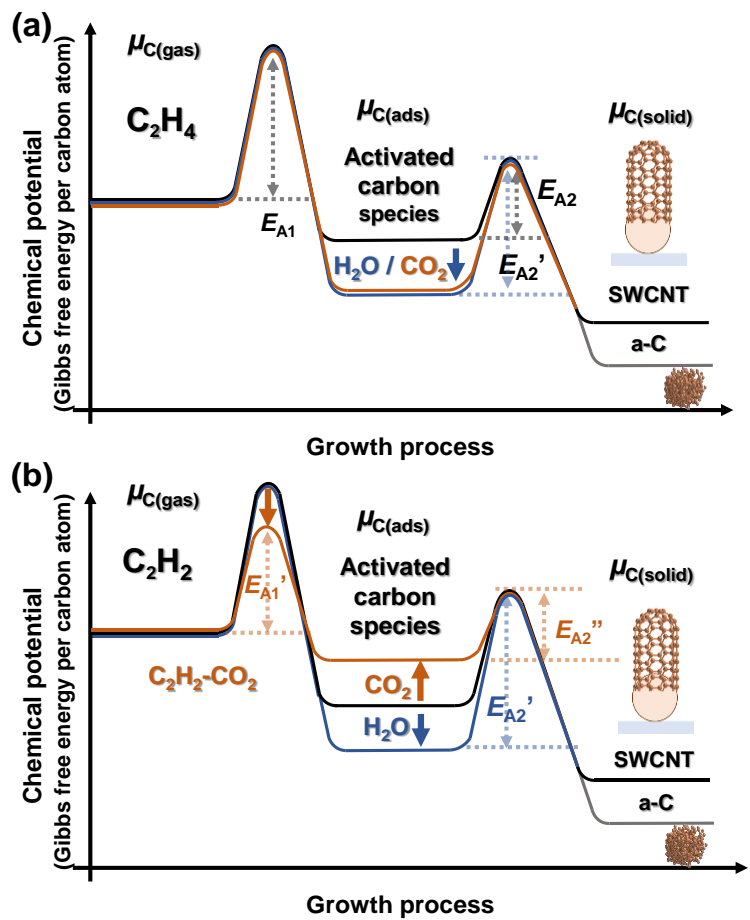

**Figure 8:** Schematic energy diagram of the SWCNT formation process from the CNP growth seeds at a high temperature using (a) $C_2H_4$ and (b) $C_2H_2$ as carbon feedstocks with different growth

enhancers. Chemical potential variation of carbon in the gas phase, activated carbon species adsorbed on the growth seed, and carbon in an SWCNT is shown. The black, blue, and orange curves denote the chemical potential under the growth conditions without growth enhancers, with $H_2O$, and with $CO_2$, respectively.

## 4. Conclusion

In conclusion, by employing ND-derived CNPs as the growth seeds, we investigated the different roles of growth enhancers ($CO_2$ and $H_2O$) in high-temperature SWCNT growth processes using different carbon feedstocks ($C_2H_2$ and $C_2H_4$). By comparing two commonly used hydrocarbons, we found that the enhancement role of $CO_2$ differed depending on the carbon source, while similar etching role of $H_2O$ exhibited in both hydrocarbon-supplied growth. In the $C_2H_4$-supplied growth, controlling the partial pressure and injection time of $CO_2$ prevents the formation of a-C without causing the formation of defects and achieves highly crystalline SWCNTs with high purity. However, $CO_2$ exhibits the enhancement effect in both SWCNT growth and deposition of a-C when using $C_2H_2$ as the carbon source. Moreover, in the two-step growth process, the injection of $CO_2$ results in a prevention effect of a-C deposition in the $C_2H_4$-supplied case and the enhancing effect of SWCNT yield and a-C deposition in the $C_2H_2$ case. Particularly in the $C_2H_4$-supplied two-step SWCNT growth process, with the use of $CO_2$, highly crystalline SWCNTs with less a-C deposition were achieved when the temperature of the secondary growth step was 1000 °C. Furthermore, by tuning the growth condition, the SWCNT yield was highly increased as indicated by the GOF of 100% and the $I_G/I_{Si}$ ratio of ~0.15, while the quality of SWCNTs was preserved. Our results demonstrate the different synergetic effects of growth enhancers and carbon sources in high-temperature SWCNT synthesis based on the solid carbon growth seeds, proving that $CO_2$ can be used as an effective growth enhancer in the $C_2H_4$-supplied two-step growth process to achieve high-quality SWCNTs with a high yield.

**Associated Content**

**Author Information**

**Corresponding Author**

**Yoshihiro Kobayashi** – *Department of Applied Physics, Osaka University, Suita, Osaka 565-0871, Japan; Email: kobayashi@ap.eng.osaka-u.ac.jp; Tel: +81 06 6879 7833; Fax: +81 06 6879 7863*

**Mengyue Wang** – *Department of Applied Physics, Osaka University, Suita, Osaka 565-0871, Japan;*

**Authors**

Yuanjia Liu – *Department of Applied Physics, Osaka University, Suita, Osaka 565-0871, Japan;*

Manaka Maekawa – *Department of Applied Physics, Osaka University, Suita, Osaka 565-0871, Japan;*

Michiharu Arifuku – *Nippon Kayaku Co., Ltd., 31-12, Shimo 3-chome, Kita-ku, Tokyo 115-8588, Japan;*

Noriko Kiyoyanagi – *Nippon Kayaku Co., Ltd., 31-12, Shimo 3-chome, Kita-ku, Tokyo 115-8588, Japan;*

Taiki Inoue – *Department of Applied Physics, Osaka University, Suita, Osaka 565-0871, Japan;*

**Notes**

The authors declare no competing financial interest.

**Acknowledgement**

The authors would like to thank Dr. T. Sakata of Research Center for Ultra-High Voltage Electron Microscopy, Osaka University, for assistance in the SEM observations. The part of this work was finantially supported by JSPS KAKENHI (Grant Numbers JP15H05867 and JP17H02745). This work was also supported by Nanotechnology Platform of MEXT (Grant Number JPMXP09F20OS0002).

# Supporting information

# Combination effect of carbon source and growth enhancer on single-walled carbon nanotube synthesis from solid carbon growth seeds

*Mengyue Wang, * [a] Yuanjia Liu, [a] Manaka Maekawa, [a] Michiharu Arifuku, [b] Noriko Kiyoyanagi, [b] Taiki Inoue, [a] Yoshihiro Kobayashi*[a]*

[a] *Graduate school of engineering, Osaka University, Suita, Osaka 565-0871, Japan*

[b] *Nippon Kayaku Co., Ltd., 31-12, Shimo 3-chome, Kita-ku, Tokyo 115-8588, Japan*

*Email: wang.my@ap.eng.osaka-u.ac.jp, kobayashi@ap.eng.osaka-u.ac.jp

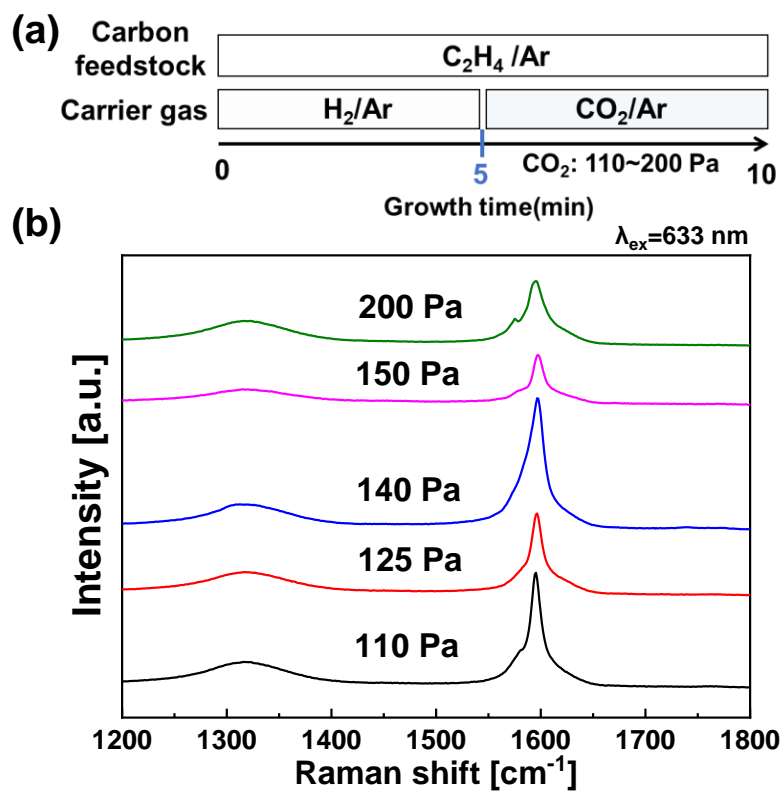

**Figure S1:** (a) Gas composition profile during one-step SWCNT growth process at 1000 °C for 10 min. In the $C_2H_4$-supplied case, $CO_2$ was injected at 5 min after SWCNT growth started, and the partial pressure of $CO_2$ was varied from 110 Pa to 200 Pa. (b) Raman spectra of SWCNTs obtained by one-step growth from $C_2H_4$ with the different partial pressure of $CO_2$ injected at 5 min after the growth starts.

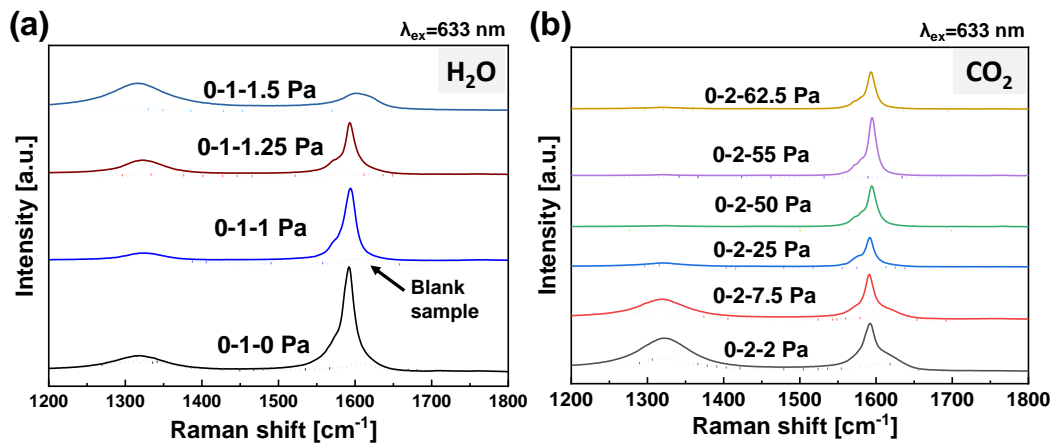

**Figure S2:** (a) Raman spectra of SWCNT samples synthesized by $C_2H_4$-supplied two-step growth with the injection of $H_2O$. The partial pressure of $H_2O$ was fixed to 1 Pa in the temperature and varied from 0 Pa to 1.5 Pa (0-1-0 Pa, 0-1-1 Pa, 0-1-1.25 Pa and 0-1-1.5 Pa) in the secondary growth step. (b) Raman spectra of SWCNT samples synthesized by two-step growth with the injection of $CO_2$. The partial pressure of $CO_2$ was fixed to 2 Pa in the temperature and varied from 2 Pa to 62.5 Pa (0-2-2 Pa, 0-2-7.5 Pa, 0-2-25 Pa, 0-2-50 Pa, 0-2-55 Pa, and 0-2-62.5 Pa) in the secondary growth step.

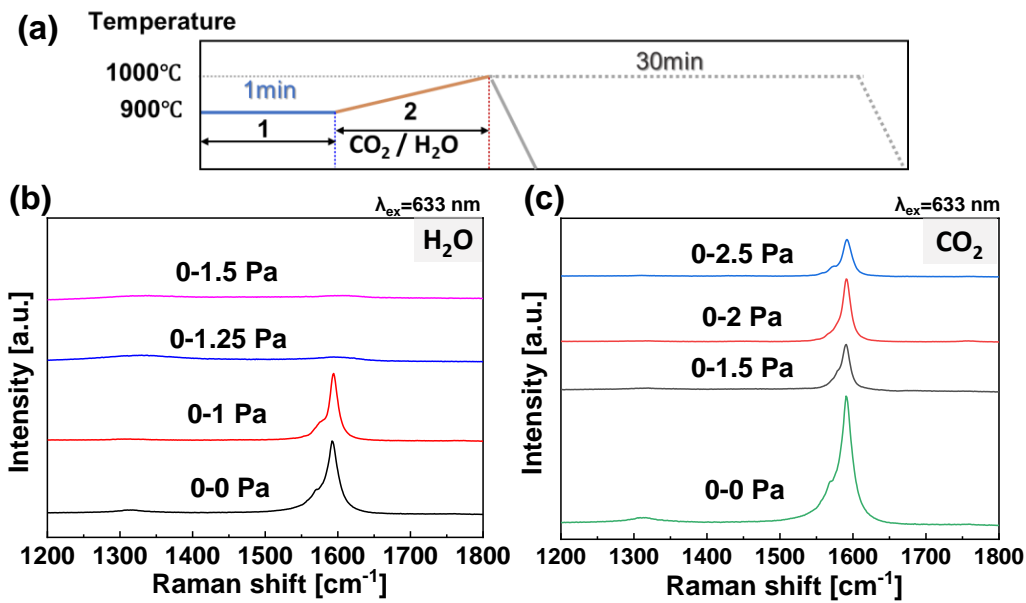

**Figure S3:** (a) Profiles of temperature and growth enhancer partial pressure for the SWCNT two-step growth process with growth enhancer injection. Adjustment of the growth enhancers partial pressure in the temperature rising step (process 2) in the $C_2H_4$-supplied two-step growth process. The SWCNT samples were collected after the finish of the temperature rising step. (b) Raman spectra of the SWCNT samples grow with the different partial pressure of $H_2O$ (from 0 Pa to 1.5 Pa) in the temperature rising step. (c) Raman spectra of the SWCNT samples grow with the different partial pressure of $CO_2$ (from 0 Pa to 2.5 Pa) in the temperature rising step.

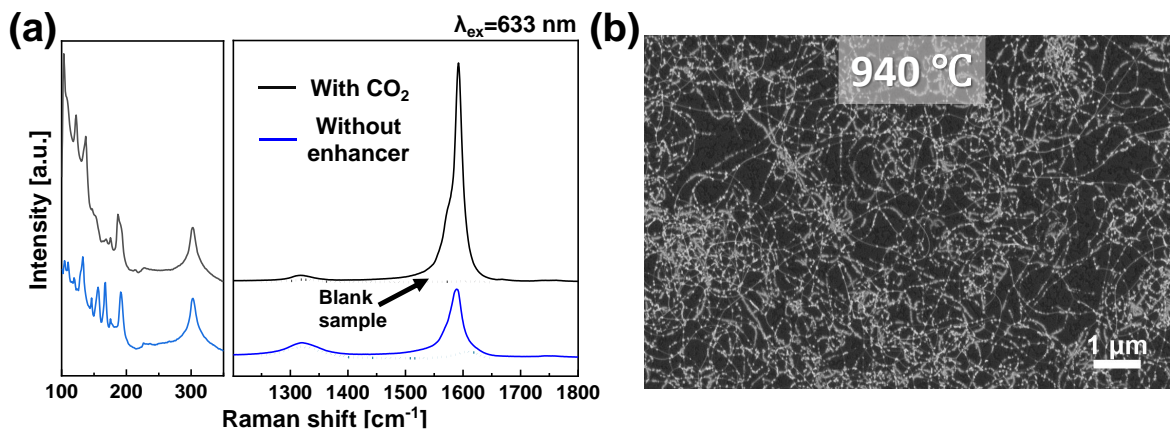

**Figure S4:** (a) Raman spectra of SWCNT samples synthesized by $C_2H_2$-supplied two-step growth with and without the injection of $CO_2$. During the growth, the growth temperature was 900 °C in the initial growth step and the temperature rise to 940 °C in the secondary growth step. The partial pressure of $C_2H_4$ was 20 Pa in the initial growth step, then changed to 5 Pa in the temperature rising step, and varied to 10 Pa when the secondary growth step started. The partial pressure of $CO_2$ was 2 Pa in the temperature and varied to 400 Pa in the secondary growth step (0-2-400 Pa). (b) SEM image of SWCNTs synthesized with the injection of CO2 (0-2-400 Pa).

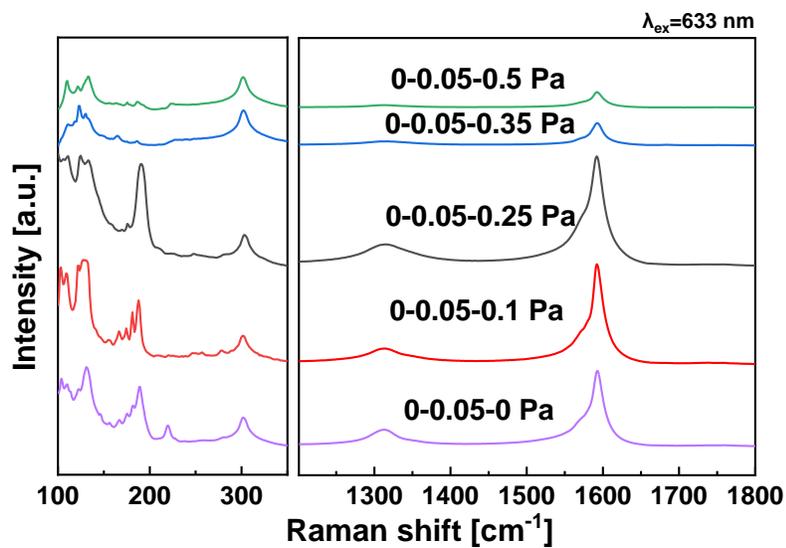

**Figure S5:** Raman spectra of SWCNT samples synthesized by $C_2H_2$ -supplied two-step growth with the injection of $CO_2$. The partial pressure of $CO_2$ was fixed to 0.05 Pa in the temperature and varied from 0 Pa to 0.5 Pa (0-0.05-0 Pa, 0-0.05-0.1 Pa, 0-0.05-0.25 Pa, 0-0.05-0.35 Pa, and 0-0.05-0.5 Pa) in the secondary growth step.

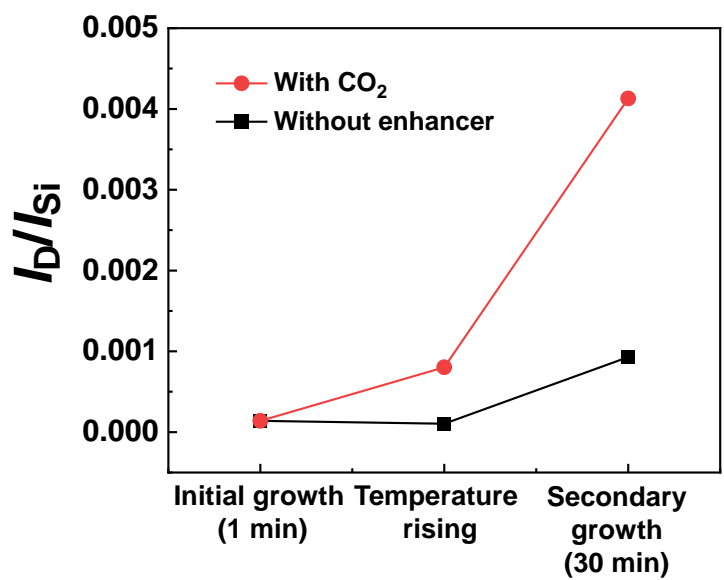

**Figure S6:** Changes in a-C deposition ($I_D/I_{Si}$ ratio) of blank sample along the growth steps in $C_2H_2$-supplied process without growth enhancers and with $CO_2$ (0-0.05-0.1 Pa).